\documentclass[sigconf]{acmart}

\usepackage{hyperref}
\hypersetup{
    colorlinks=true,
    linkcolor=blue,
    filecolor=magenta,      
    urlcolor=cyan,
}
 
\AtBeginDocument{%
  \providecommand\BibTeX{{%
    \normalfont B\kern-0.5em{\scshape i\kern-0.25em b}\kern-0.8em\TeX}}}

\makeatletter
\newcommand\addauthornote[1]{%
  \if@ACM@anonymous\else
    \g@addto@macro\addresses{\@addauthornotemark{#1}}%
  \fi}
\newcommand\@addauthornotemark[1]{\let\@tmpcnta\c@footnote
   \setcounter{footnote}{#1}\addtocounter{footnote}{-1}
    \g@addto@macro\@currentauthors{\footnotemark\relax\let\c@footnote\@tmpcnta}}
\makeatother

\newcommand{\ie}{\emph{i.e.},~}
\newcommand{\eg}{\emph{e.g.},~}

\def\F{Figure~}
\def\T{Table~}

\setcopyright{acmcopyright}
\copyrightyear{2020}
\acmYear{2020}
\acmDOI{10.1145/1122445.1122456}

\begin{document}

\title{A Machine Learning System for Retaining Patients in HIV Care}


\author{Avishek Kumar}
\authornote{These authors contributed equally towards this work.}
\authornote{This work was done while the author was at the University of Chicago}
\affiliation{%
  \institution{Intuit AI}
}
\email{avishek.raman.kumar@gmail.com}

\author{Arthi Ramachandran}
\addauthornote{0}
\addauthornote{1}
\affiliation{%
  \institution{Civis Analytics}
}
\email{arthir@cs.columbia.edu}

\author{Adolfo De Unanue} 
\addauthornote{2}
\affiliation{%
  \institution{ITAM}
}
\email{unanue@itam.mx}

\author{Christina Sung} 
\affiliation{%
  \institution{University of Chicago}
}
\email{christina.sung@gmail.com}

\author{Joe Walsh}
\affiliation{%
  \institution{University of Chicago}
}
\email{jtwalsh@protonmail.com}

\author{John Schneider}
\affiliation{%
   \institution{University of Chicago Medicine}
}
\email{jschnei1@medicine.bsd.uchicago.edu}

\author{Jessica Ridgway}
\affiliation{%
   \institution{University of Chicago Medicine}
}
\email{Jessica.Ridgway@uchospitals.edu}

\author{Stephanie Masiello Schuette}
\affiliation{%
   \institution{Chicago Department of Public Health}
}
\email{Stephanie.Schuette@cityofchicago.org}

\author{Jeff Lauritsen}
\affiliation{%
   \institution{Chicago Department of Public Health}
}
\email{Jeff.Lauritsen@cityofchicago.org}

\author{Rayid Ghani}
\addauthornote{2}
\affiliation{%
  \institution{Carnegie Mellon University}
}
\email{rayid@cmu.edu}

\thanks{

Any opinions, findings, and conclusions or recommendations expressed in this material are those of the authors and do not necessarily reflect the views of our employers.}

\graphicspath{ {figures/} }

\renewcommand{\shortauthors}{Kumar and Ramachandran, et al.}

\begin{abstract}
Retaining persons living with HIV (PLWH) in medical care is paramount to preventing new transmissions of the virus and allowing PLWH to live normal and healthy lifespans. Maintaining regular appointments with an HIV provider and taking medication daily for a lifetime is exceedingly difficult. 51\% of PLWH are non-adherent with their medications and eventually drop out of medical care.  Current methods of re-linking individuals to care are reactive (after a patient has dropped-out) and hence not very effective.  We describe our system to predict who is most at risk to drop-out-of-care for use by the University of Chicago HIV clinic and the Chicago Department of Public Health. Models were selected based on their predictive performance under resource constraints, stability over time, as well as fairness. Our system is applicable as a point-of-care system in a clinical setting as well as a batch prediction system to support regular interventions at the city level. Our model performs 3x better than the baseline for the clinical model and 2.3x better than baseline for the city-wide model. The code has been released on github \footnote{\url{github.com/dssg/hiv-retention-public}} and we hope this methodology, particularly our focus on fairness, will be adopted by other clinics and public health agencies in order to curb the HIV epidemic.
\end{abstract}

\ccsdesc[500]{Applied computing~Health informatics}
\ccsdesc[300]{Applied computing~Health care information systems}
\ccsdesc[300]{Computing methodologies~Machine learning}

\keywords{Public Health, Machine Leaning, Social Good, HIV}

\maketitle

\section{Introduction}
HIV has become one of the most devastating global pandemics in modern history, infecting over 75 million people on all parts of the planet and leading to 32 million deaths globally\cite{unaids_2019}. In the City of Chicago HIV diagnoses are 1.5x the national rate and only 36\% of the HIV population is retained in medical care\cite{cdph_chicago_2018}.  The state-of-the-art HIV treatment is antiretroviral therapy (ART). HIV-positive individuals who are retained in care and taking antiretroviral therapy are able to suppress their HIV viral level in their serum to undetectable levels, effectively eliminating the risk of transmitting HIV to others and live normal lifespans~\cite{gardner_spectrum_2011, hiv_uu_2019}. For ART to be effective, a person must take medication everyday and regularly see a doctor for the entirety of their life. If the majority of persons living with HIV (PLWH) were virally suppressed through ART, it is possible to have functionally zero new HIV infections and end the HIV epidemic. The problem of combating HIV no longer solely lies in developing effective treatment. Now, the problem lies in keeping PLWH retained in care and virally suppressed for their lifetime. In response, there has been a call for new ideas and interventions to scale-up participation in the HIV care continuum to achieve universal viral suppression. 

The HIV care continuum describes the stages of care necessary to achieve viral suppression: 1) diagnosis of HIV; 2) linkage-to-care, link to medical care and prescription of ART; 3) retention in care, attending medical appointments on a regular basis; and 4) viral suppression, no detectable HIV\cite{cdc_care_continuum}. Retention in care is not only important for the individual health of people living with HIV, but also for public health. Accordingly, retention and accessing care is a critical pillar of public health agency plans to eliminate HIV transmission in the United States. 

However, in the U.S., less than half of individuals living with HIV are retained in care. The causes for the low retention are multi-faceted. There are several state and federal programs, such as the Ryan White HIV/AIDS program, to provide funding for HIV care visits and medications. Despite these programs, many PLWH still do not regularly attend medical appointments. Many social, economic, and personal factors play a role in retention in care, including mental illness, substance use, insecure housing, poverty, neighborhood violence, and stigma~\cite{ cunningham_factors_2014, giordano_predictors_2009, almirol_hiv_2016, giordano_patients_2005, cook_illicit_2007, zuniga_role_2016}. Effective interventions take personalized approaches and include intensive case management, peer navigation, and multi-faceted outreach programs~\cite{horstmann_retaining_2010, okeke_enhancing_2014, gardner_enhanced_2014, higa_interventions_2012}. 

While these interventions are effective, they are also  resource intensive. Further, most clinics and public health settings have limited resources at their disposal.
It is estimated that 86\% of PLWH in the U.S. are diagnosed as of 2015~\cite{pecoraro_factors_2013}. However, only 49\% of them are retained in care and 51\% are virally suppressed\cite{cdc_care_continuum}. Therefore, methods are needed to identify and prioritize PLWH who are at the highest risk for falling out of medical care in order to prioritize their needs of being kept in care.

Existing work on this problem has focused on two aspects: 1) using retrospective analysis to identify coarse, population level subgroups at risk for dropping out of care, such as African-American men who have sex with other men~\cite{mayer_concomitant_2014, geng_retention_2010}, and 2) understanding root causes and barriers to retention in care, such as mental illness, substance use, insufficient means of transportation, lack of insurance, homelessness, disruption of social and sexual networks, unemployment, and neighborhood characteristics\cite{aidala_homelessness_2005,baillargeon_accessing_2009, grinstead_reznick_effectiveness_2011, hedrich_effectiveness_2012, cooper_risk_2016, mcfadden_dynamic_2014, mayer_concomitant_2014}. These approaches are not actionable because while retrospective analysis to find subgroups is useful in describing the at-risk groups, it is not useful in proactively targeting resources. Targeting interventions using coarse, group level risk factors (e.g., men who have sex with other men) waste scarce resources because it presumes that all members have uniform risk, neglecting individual circumstances and behaviors. In contrast, a more fine-grained machine learning approach can overcome these shortcomings. 

\subsection{Our Contribution}
In this paper, we describe an HIV-specific machine learning platform built in collaboration with the Chicago Department of Public Health and the University of Chicago HIV Clinic that increases engagement in the HIV continuum by shifting from reactive interventions--patient has dropped out of care, possibly locating, and re-linking to care--to proactive interventions. Improved targeting of at-risk individuals can reduce the incidence and prevalence of HIV by keeping more PLWH individuals in-care and virally suppressed, eliminating HIV-transmission channels. The platform explores the use of administrative, surveillance, electronic medical records, and domain knowledge from HIV experts through machine learning models that are scalable, adaptive, and produce patient-level predictions with associated risk factors for each prediction for proactive intervention. \textit{The system is designed to optimize and balance i) different resource constraints of different settings, ii) stability in performance over time, and iii) fairness to prevent biases in protected groups.} The system is designed to support interventions both when a patient is at a clinic (at the time of an appointment) as well as routine proactive outreach by a public health department, informing everyday treatment decisions (e.g., which clinic is best suited for an individual) and policy decisions (e.g., what types of programs lead to a successful intervention).

\section{HIV Retention Problem}
This work was done with two partners: The University of Chicago Medicine HIV clinic (UCM) and the Chicago Department of Public Health (CDPH) to 1) support two different use cases and deployment scenarios and 2) test the effectiveness on two different data sets.

\subsection{HIV Clinic}
\begin{figure}
	\includegraphics[width=\linewidth]{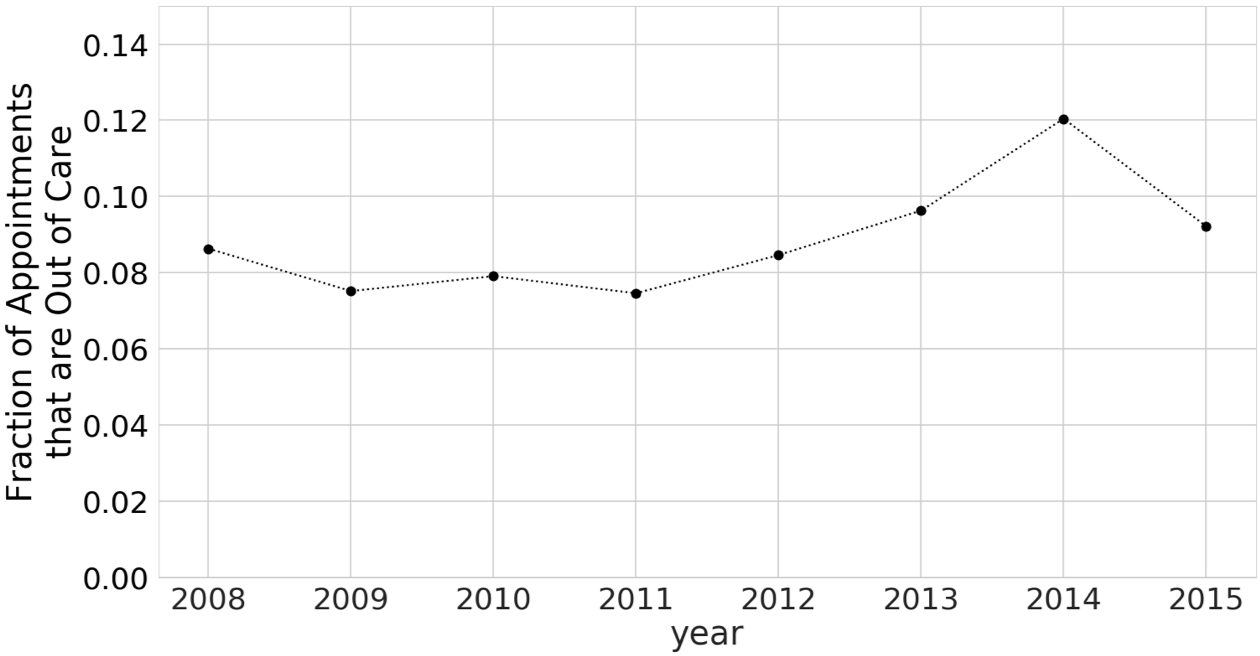}
    \caption{Approximately 10\% of all appointments at the UCM HIV clinic have no follow-up.}
    \label{FIG:hivucm_fig1}
\vspace*{-\baselineskip}

\end{figure}
The University of Chicago Medicine HIV clinic (UCM) has a predominantly reactive approach to re-linking patients where they attempt to contact patients via phone if they have not been seen in the last 12 months. Over the study period, patients attended at least one HIV care appointment, accounting for a  total of 1200 visits. Of these appointments, between 8-12\% of appointments were not followed by a subsequent appointment at least 90 days apart in a  12-month period, indicating a lack of retention in care for that time period (\F\ref{FIG:hivucm_fig1}). Using another measure of engagement, access to care, approximately 10\% of the appointments did not have a subsequent appointment in a six month period, meaning that patient did not access care within six months. Our system seeks to identify these patients before they drop out of care and prioritize them to keep them in medical care.

\subsection{Department of Public Health}
The Chicago Department of Public Health (CDPH) also currently takes a reactive approach to re-linking PLWH back into care by identifying a set of people who have already been been out of medical care for approximately 18 months. On average 28\% of patients do not access HIV care within a 12 month period, approximately 4000 persons. A list of PLWH who have not been in care is then provided to bridge-to-care workers who do outreach to locate and re-link persons to care on a monthly basis. The scope of the current work is to provide lists of PLWH at high risk of dropping-out-of-care on a monthly basis \textit{before} they drop out of care to the bridge-to-care workers using the data that is already being collected by CDPH.

\section{Data Sources}
\subsection{HIV Clinic Data}
The clinic cohort includes HIV-positive individuals 18 years of age and older who attended at least one medical appointment at the University of Chicago adult HIV care clinic between  January 1, 2008 and May 31, 2015. For all eligible patients, the following data is available from the EMR (Electronic Medical Record): demographics, appointment history, insurance information, other medical conditions, medications, HIV care provider, substance use history, and laboratory test results. Laboratory test results collected included HIV viral load, lymphocyte subset data (e.g., CD4 count), sexually transmitted infection (STI) test results, and toxicology test results. \T\ref{tab:table_ucmdata} provides a list of the types of the data provided by the UChicago HIV Clinic.

\begin{table}[h]
  \begin{center}
    \begin{tabular}{ |p{2cm}|p{6cm}| }
    \hline
      \textbf{Data Type}  & \textbf{Fields}\\ 
      \hline
      Demographics  & Age, Gender, Race, Address\\ \hline 
      Lab Tests  & CD4, Viral Load\\ \hline 
      Appointment History & Number/Date of Appointments 
                            with all providers \\ \hline
      Diagnoses  & Psychiatric Illness, Opportunistic Infection, STD, Substance Abuse\\ \hline 
      Medications  & All Medications Prescribed\\ \hline 
    \end{tabular}
    \caption{Data provided by UChicago HIV Clinic}
    \label{tab:table_ucmdata}
  \end{center}
\end{table}
\vspace*{-\baselineskip}
Patients’ addresses are geocoded and the travel distance and travel time to the clinic as well as the crime rate along the travel route are calculated  \cite{ridgway_travel_2018}. Using data from the American Community Survey (US Census Bureau), characteristics of a patient’s community at the census tract level are collected, specifically, the racial composition, fraction of population on Supplemental Nutrition Assistance Program, commute characteristics, and education levels \cite{census}.

\subsection{Chicago Department of Public Health Data}

\begin{table}[h]
  \begin{center}
    \begin{tabular}{ |p{2cm}|p{6cm}| }
    \hline
      \textbf{Data Type} & \textbf{Fields}\\ \hline
      Demographics  & Age, Gender, Race, Zip Code\\ \hline 
      Lab Tests  & CD4 Count, Viral Load\\ \hline 
      Opportunistic Infections   & Infection Type (e.g., Pneumonia, Tuberculosis, Cancer)\\ \hline 
      Transmission Risk  & How a person was infected with HIV (e.g., Perinatal, Sexual Contact, Percutaneous)\\ \hline
    \end{tabular}
    \caption{List of Data provided by CDPH}
    \label{tab:table_cdphdata}
  \end{center}
\end{table}

The data used for the city-wide model is data typically used for what is called surveillance in epidemiology of the PLWH individuals in the city of Chicago by the Chicago Department of Public Health. In order to protect the identity of the patients, the data used for building and testing models is deidentified: Names were removed, location was used at the granularity of zip code, and dates are shifted by a predefined number of days. \textbf{Results are reported based on the shifted dates}. 

The cohort that is evaluated are PLWH within the CDPH jurisdiction between the years 2041-2046 (dates were shifted to preserve anonymity) that currently live in the city of Chicago and have had a CD4 or Viral Load test in the last 12 months. \T\ref{tab:table_cdphdata} provides a summarized list of the type of data provided by CDPH. 

Demographic data include age, gender, race, and zip code, and HIV transmission category (\eg Perinatal, Sexual Contact, Percutaneous). Lab test data contains viral load tests, CD4 level tests, opportunistic infections and the date of each test. In HIV care, a viral load test measures the amount of virus present in a person’s blood serum. An HIV patient with a viral load below 200 copies of the virus per ml is considered virally suppressed, unable to spread the virus to others and able to live a normal lifespan, and is therefore the goal of HIV care. A CD4 count is a blood test to measure the level of CD4 cells in the body. CD4 cells are a type of white blood cell that are important in the immune system. A person with a CD4 count lower than 200 cells/mm$^3$ is considered immunosuppressed and given a diagnosis of AIDS. 

\subsubsection{Data Limitations}
The data that CDPH collects has several limitations that is typical of what is available to public health departments. EMR data regarding patients’ diagnoses, medications, etc. may be inaccurate if providers do not accurately document and update patient data at each visit. Prior studies have shown wide variability in accuracy of billing diagnoses and incomplete problem list documentation in the EMR. We attempted to limit inaccuracy due to poor documentation by incorporating multiple fields from the EMR. For example, patients with a history of substance abuse were detected not only by examining billing diagnoses for substance abuse, but also by collecting clinician-assigned diagnoses in the problem list, social history documentation of substance abuse, and toxicology screen results. Furthermore, certain factors that may have an important impact on retention in care may not be captured within structured fields of the EMR, i.e., life stressors, social support, child care or other responsibilities, etc. In the future, we plan to incorporate natural language processing of unstructured clinical notes and case works interviews into the model to detect these factors.

\section{Methodology}
We use our open-source machine learning toolkit, Triage\cite{triage_dsapp} to build this system. Triage allows for rapid and iterative creation of end to end machine learning systems. Aequitas~\cite{saleiro_aequitas:_2018}, a bias and fairness audit tool to inspect results of machine learning models for bias in order to make informed and equitable decisions, was used to measure fairness in models. 

\subsection{Feature Engineering}
 Feature design was guided by prior literature and domain expertise of HIV retention of our team. Factors previously shown to be associated with retention in HIV care are age, CD4 count, substance use, psychiatric illness, and history of prior visits. For each feature, measures were aggregated by time (e.g., number in the past six months, mean for the past year, etc.) or space (e.g., the number of assaults in the patient’s residential census tract in the past year). A range of values for time (6 months, 1 years, 3 years, all history) and space (by zip code and census tract) aggregations as well as different aggregation functions (mean, min, max, standard deviation) were calculated for each feature. Categorical variables (such as race, transmission category) were dummified using one-hot encoding. Missing data was imputed with the imputation method depending on the variable missing (e.g., a missing birth date resulted in an age assignment of the mean age of the population). An indicator flag for whether a feature was imputed was used as an additional feature, allowing the model to use missingness itself as a feature.

The HIV Clinic model used categories of features including demographics, diagnoses, location-based features, laboratory test results, medical visits, and specific providers seen resulting in 800 features. The Department of Public Health model used categories of features include demographics, lab tests, opportunistic infections, transmission category, and diagnosis status, resulting in \~200 features. 

\subsection{Retention and Access Labels}
            
We looked at predicting two different types of outcomes based on discussions with our partners at CDPH and UCM: 1) retention in care and 2) access to care. Retention in care is defined as attending at least 2 HIV care visits greater than 90 days apart within a 12-month period~\cite{mugavero_access_2010}. This definition of retention is defined by the Health Resources and Services Administration HIV/AIDS Bureau (HRSA HAB). While there is no true gold standard of retention in care, this definition has been shown to be correlated with patient health outcomes including HIV viral suppression~\cite{mugavero_measuring_2012}. It is the definition of interest to the UCM HIV clinic. Access to care is defined as having a single HIV care visit within a 6 or 12 month period~\cite{mugavero_measuring_2012, cdc_care_continuum}. As the name indicates, this label predicts whether a patient will access HIV medical care within a 6 or 12 month period. This metric is used by public health departments for the purposes of surveillance~\cite{cdph_chicago_2018}. The risk score predicted by the model can be used to inform and prioritize interventions to improve retention and access to care.

\section{Modeling Approach and Results}
The problem of identifying HIV patients at risk of dropping out of care was cast as a binary classification problem using a variety of labels that are of interest to CDPH and UCM.  Training and test sets are created for every month (public health model) or year (clinic model) in order to mimic the business process of the clinic and public health department, respectively. A variety of classification methods over a hyperparamter grid (Decision Trees, Logistic Regression, Random Forest, Gradient Boosted Decision Trees) were used to develop models before performing model selection.

\subsubsection{Temporal Cross Validation} Model selection was performed using temporal cross validation. Temporal cross validation was used instead of k-fold cross-validation to account for serial correlation and temporal effects in the data and accurately model the business process in deployment. \textit{Temporal cross validation also allows us to assess model stability over time which is not possible in conventional $k$-fold cross validation.} The data were divided into training and test sets split by time. For example in the proactive outreach scenario (CDPH), if we are assessing the risk of an individual not accessing care within the next 12 months at the time we are selecting individuals to contact (1st of every month for example), then the model is trained at the beginning of every month (e.g., January 1, 2018) using all the information available from the past data (the training set). The model can then predict on all individuals in the cohort for that month (the test set). This mimics how the model will be used in deployment and prevents temporally leaking information.

\subsubsection{Model Selection} Model performance was evaluated using precision with a population threshold based on the resource constraints of the setting. \textit{Rather than optimizing the model for AUC,  an aggregate metric, models selection is done by locally optimizing the precision-recall space to tune the model to the resource constraints of the deployment setting, providing an accurate measure of performance of the deployed model.}

In order to use retention resources efficiently, the system needs to minimize false positives, which minimize wasting resources on patients who will not drop-out-of-care.  To prioritize a small number of individuals for intervention, precision for the top $k$\%, where $k$ is determined by resource constraints, ensures the model selected will minimize false positives within the intervention set. The final model selected was chosen for having consistently high performance over the last five time periods in order to ensure consistent performance over time. Specifically, the model that most frequently was within 5\% of the precision of the best possible model over the last fives time periods was chosen (e.g., if the best possible precision for a time period was 0.7, all models above 0.65 precision were selected). This method of selection ensures both stability and performance in the final model deployed to the clinic or health department.

\subsection{HIV Clinic Model}
The clinic model is designed to make a prediction at the time of each patient’s HIV care appointment, replicating the workflow (and data available) in the clinic, where the patient arrives for their appointment and then receives a risk score. An intervention can then be initiated during their appointment. In the case of the HIV clinic, models were selected based on precision for the top 10\% of risk scores. This value was chosen based on the intervention capacity of the HIV clinic. If the model were to be adopted by other clinics, the threshold can be adjusted to meet the resources of that clinic. 

In the cohort used to validate the model,  patients had to have attended at least one HIV care appointment. Of these appointments, between 8-12\% of appointments were not followed by a subsequent appointment at least 90 days later within a 12-month period, indicating a lack of retention in care for that time period (Figure \ref{FIG:hivucm_fig1}). Also, of these appointments, approximately 11\% of the appointments did not have a subsequent appointment in a six month period (access to care).

\subsubsection{HIV Clinic Access to Care Model}
\begin{figure}[h]
	\includegraphics[width=\linewidth]{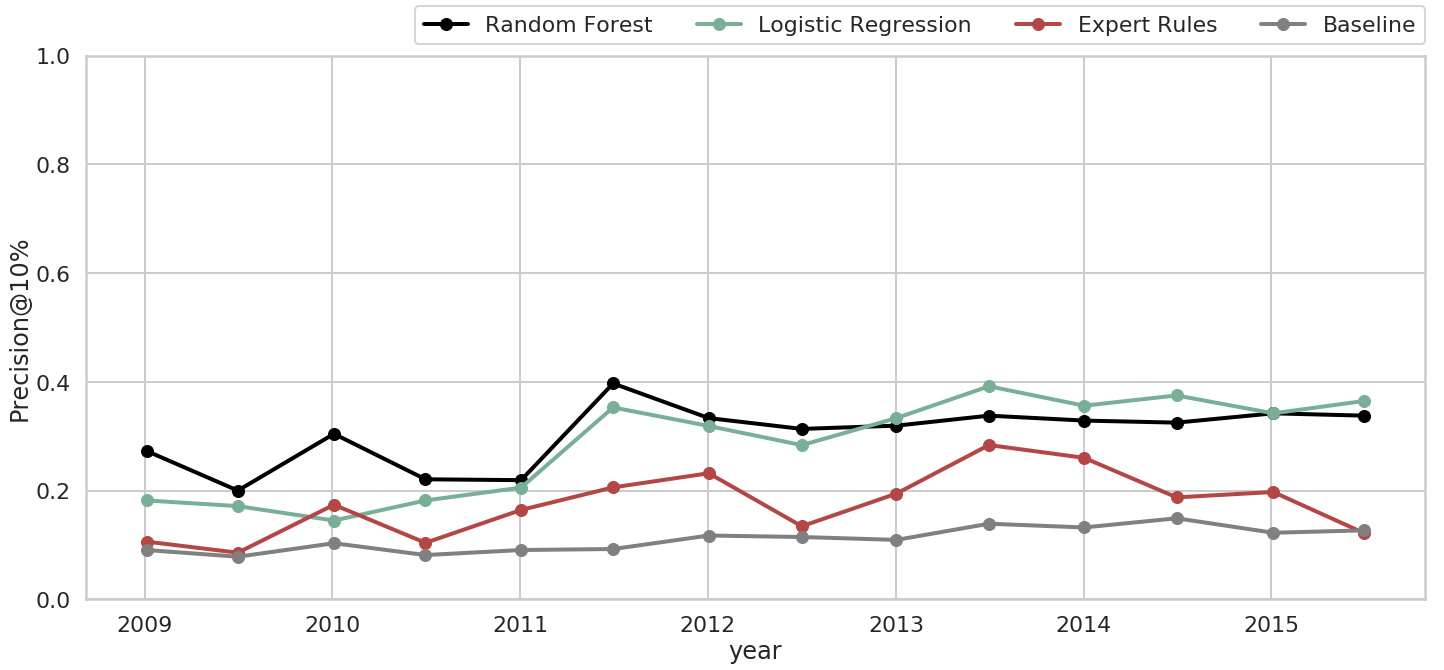}
    \caption{Precision@10 of UCM Access to Care. The best performing model is a Random Forest based on its precision@10\% and the stability of results over the last five time periods. The Random Forest model is 3x better than the baseline and 1.7x better then Expert Rules given by domain experts. }
    \label{FIG:ucm_access_precision_time}
\end{figure}

\begin{table}
\begin{tabular}{|l|c|c|}
\hline
\textbf{Model} & \textbf{Precision@10\%}  & \textbf{Number of}    \\ 
& &  \textbf{Appointments}    \\ 
& &  \textbf{Correctly Flagged}    \\ \hline
Random Forest & 0.30 $\pm$ 0.05   & 21  \\ \hline
Logistic Regression & 0.29 $\pm$ 0.09 & 21 \\ \hline
Expert Rules & 0.18 $\pm$ 0.06 & 13  \\ \hline
Baseline & 0.11 $\pm$ 0.02 & 8 \\ \hline
\end{tabular}
\caption{Model Performance of UCM Access to Care Model}
\label{tab:ucm_access_performance}
\end{table}
\vspace*{-\baselineskip}

The HIV Clinic Access to Care Model predicts the risk of a patient not accessing care within six months. Models were evaluated based on stability over time and Precision@10\%. The best performing model for predicting access to care is a Random Forest (1000 trees, 10 min samples/leaf, no max depth). The Random Forest model model is 3x better than the baseline, correctly flagging approximately 3x more appointments than the baseline (prior) and 1.7x better then Expert Rules provided by domain experts at the clinic. The Expert Rules are based on age, length of time on ART, substance abuse and viral suppression. \F\ref{FIG:ucm_access_precision_time} shows a Logistic Regression model has similar performance to the Random Forest but is less consistent over time; therefore, the Random Forest model was selected.  The performance of the expert rules is highly inconsistent over time and the performance decays over time, indicating the rules are not taking into account a shift in the data over time. The most predictive features of the UCM access model are appointment history and retention history, particularly the number of days between appointments and number of completed appointments. This indicates that the history of appointments as well as the cadence a patient has in accessing care is important for remaining in medical care. 

\subsubsection{UCM Retention in Care Model}
The HIV Clinic Retention in Care Model predicts the risk of a patient not be being retained in care (not having 2 appointments within 90 days in a 12 month period). This label has been shown to be correlated with effective care and is therefore of great interest to the UCM clinic. Models were evaluated based on stability over time and Precision@10\%, matching the clinic's capacity for intervention (150 appointments/year). The best performing UCM Retention model is a Random Forest (5000 estimators, max depth 5, 10 minimum samples split). The Random Forest model is 2x better than the baseline (prior) and 1.7x better than a simple Decision Tree and previously published Expert Logistic Regression Model\cite{ridgway_2018}, leading to flagging 18 and 15 more appointments, respectively. The Random Forest model is also considerably more stable over time compared to all other models. The most predictive features of the model include the consecutive days a patient has been retained, the number of days between appointments, and the number of viral load tests where a patient has been virally suppressed. These features indicate that the history of retention and viral suppression are predictive of future retention in care.   

\begin{figure}[h]
	\includegraphics[width=\linewidth]{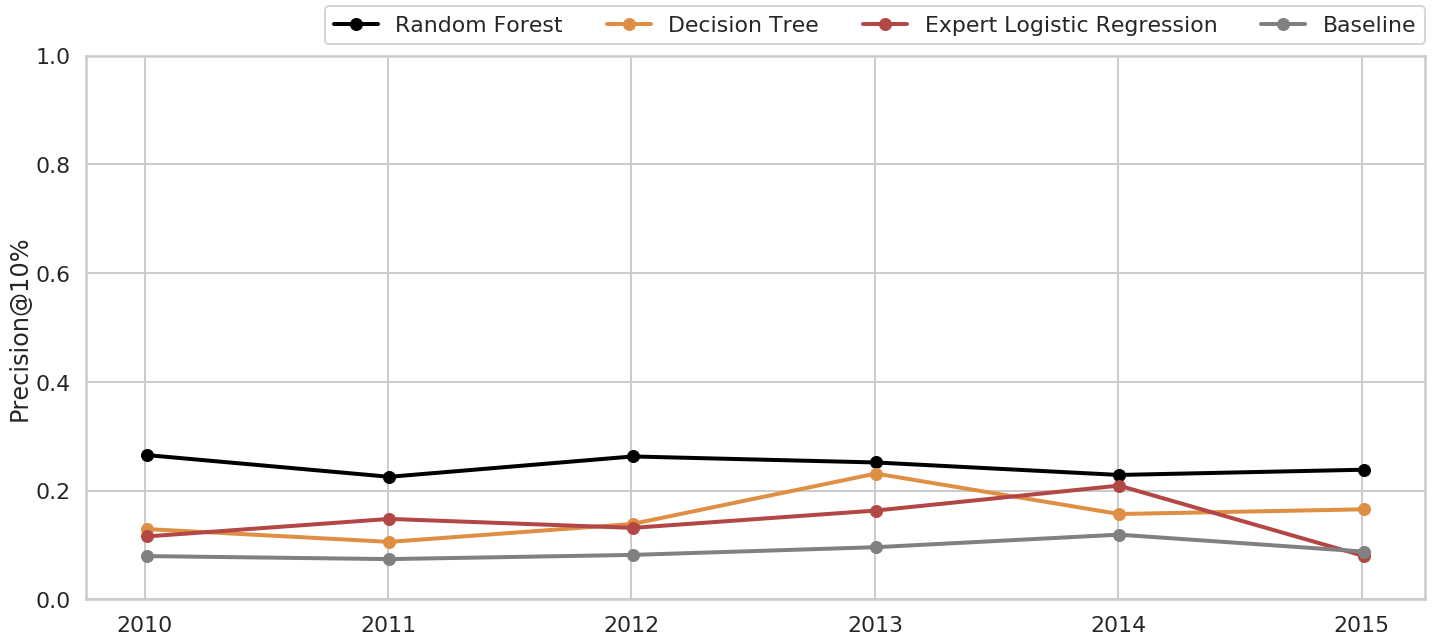}
    \caption{Precision@10\% of HIV Clinic Retention in Care Model. The best performing model is a Random Forest based on its precision@10\% and the stability of results over the last five time periods. The Random Forest model is 2x better than the baseline and 1.7x better then a simple Decision Tree and previously published Expert Logistic Regression Model. }
    \label{FIG:ucm_retention_precision_time}
\end{figure}

\begin{table}
\begin{tabular}{|l|c|c|}
\hline
\textbf{Model} & \textbf{Precision@10\%}  & \textbf{Number of}    \\ 
& &  \textbf{Appointments}    \\ 
& &  \textbf{Correctly Flagged}    \\ \hline
Random Forest & 0.25 $\pm$ 0.02   & 38  \\ \hline
Decision Tree & 0.15 $\pm$ 0.04 & 23 \\ \hline
Expert Logistic Regression & 0.14 $\pm$ 0.04 & 21  \\ \hline
Baseline & 0.13 $\pm$ 0.10 & 20 \\ \hline
\end{tabular}
\caption{Model Performance of UCM Clinic Retention in Care Model}
\label{tab:ucm_retention_performance}
\end{table}

\subsubsection{Feature Importances}
In order to sanity check the models as well as help explain to clinicians at UCM and public health experts at CDPH what signals the machine learning models were picking up on, we generated feature importances for our selected models. The models for both retention and access to care rely on similar predictor variables, sharing 80\% of the top 20 predictors. Behavioral features such as past history of retention in care and previous HIV care encounters were found to be most predictive in both access and retention models. The regression model found demographic features--race, ethnicity, days since diagnosis--to be the most predictive. The best Random Forest initially found demographic features to be important, but the models on later time periods found behavioral features to be more predictive than demographic features. 

\subsection{Health Department Retention Model}

\begin{figure}[h]
	\includegraphics[width=\linewidth]{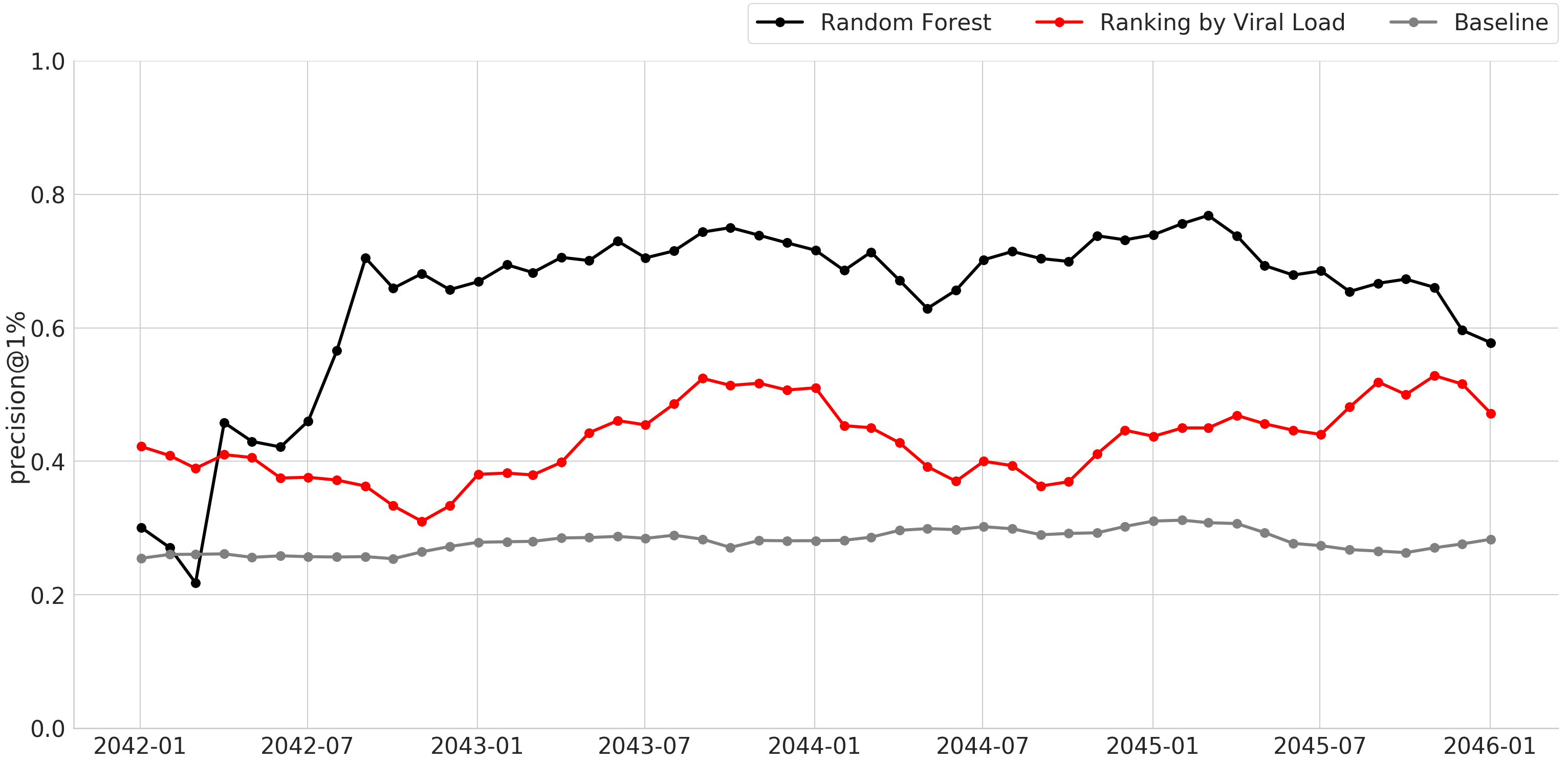}
    \caption{The Department of Public Health model for Access to Care is on average 2.5x better than the baseline and 1.5x better than simply ranking patients by their viral load.}
    \label{FIG:hivcdph_fig2}
\end{figure}

\begin{figure}[h]
	\includegraphics[width=\linewidth]{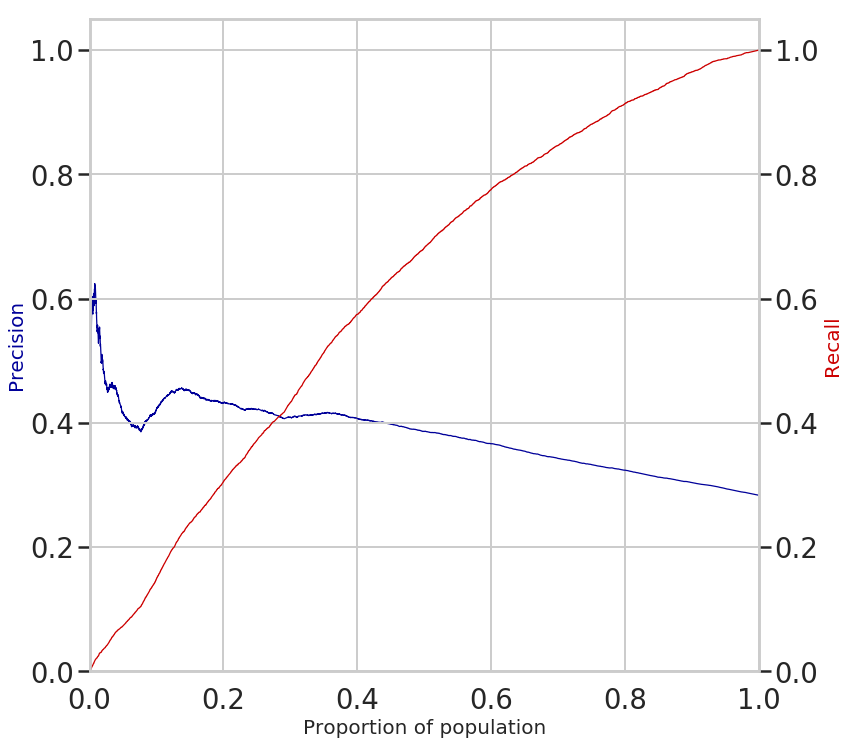}
    \caption{Precision and Recall for Last Month of Access to Care 12 months: The Precision at 1\% for the last time-split is 0.57, which results in identifying 92/161 people individuals that will not access care. This is twice as many people as the baseline rate of 45/161 people}
    \label{FIG:hivcdph_fig3}
\end{figure}

\begin{table}[h]
\begin{tabular}{|l|c|c|}
\hline
\textbf{Model} & \textbf{Precision@1\%}  & \textbf{Number of}    \\ 
& &  \textbf{People}    \\ 
& &  \textbf{Correctly Flagged}    \\ \hline
Random Forest & 0.65 $\pm$ 0.13   & 107  \\ \hline
Ranking by Viral Load & 0.43 $\pm$ 0.06 & 71 \\ \hline
Baseline & 0.28 $\pm$ 0.02 & 46 \\ \hline
\end{tabular}
\caption{Model Performance of Department of Public Health Access Care within 12 months Model}
\label{tab:cdph_access_performance}
\end{table}
CDPH is focused on flagging individuals who are at high risk of not accessing care within 12 months, \ie whether a HIV person will see a doctor within a year from the date of prediction. In the current process, CDPH will generate a list of patients, approximately 100-150 patients, that have already dropped out of care on at monthly basis, and then spend the month attempting to re-link those patients to medical care. The model is designed to mimic the current process by optimizing for precision for the top 1\% (roughly 100-150 people/month).

The specific label used by the Department of Public Health model was assessing the risk of a PLWH not accessing HIV medical care within the next 12 months from the date of prediction. The best model for access to care in 12 months is a Random Forest (1000 trees and max depth of 2) model that is 2.3x better than the baseline (prior), meaning the model is capturing 230\% more people (107 people) than simply labeling everyone as out-of-care (46 people). The best performing model was also compared to a more realistic practice, ranking patients by their viral load. Viral load measures how much virus is present in a person's serum. If a person is not virally supressed they are able to transmit the virus to others. The best model was 1.5x better than a simple viral load ranking, indicating that the viral load is not perfectly predictive of whether a patient will not access care in the next 12 months. \T\ref{tab:cdph_access_performance} summarises the performance of the Random Forest, Ranking, and Baseline. 

\F\ref{FIG:hivcdph_fig3} is the Precision and Recall for the top k\% of last year's CDPH 12 month access model. The Figure can be used as a menu for forecasting the resources needed to achieve the desired results. For instance, CDPH currently has capacity for the top 1\% of the HIV population, resulting in a precision@1 of 0.57 (flagging 92/161 people). If, say, CDPH wanted to intervene on half of all people who will not access care (recall@50\%) they would need to intervene on approximately 30\% of the population (5000 people) to achieve that goal. The precision-recall curve can provide a policy menu for understanding the type of results to expect given the amount of resources used for interventions. 

The top predictive features of the model are previous appointment history, viral load tests, and CD4 tests. Notably, demographics, transmission category, diagnosis status, and zip code were not as predictive in the model in the later years. This is typical in machine learning models where demographic variables are not as predictive as behavioral variables, especially as the amount of behavioral data increases.

\section{Bias and Fairness of our Models}
Machine learning models deployed in these two settings with many at-risk groups involved have the potential to disproportionately affect some sub-groups and exacerbate disparities. It is not sufficient to only select models based on their efficiency and effectiveness.  For instance, if the rate of dropping out of care were the same for white and black HIV patients, but the model consistently selected white PLWH for intervention, the act of using the model for intervention would create a racial disparity.  \textit{Models should be audited for biases as a part of the model selection process. A model should then be selected based on both its performance and fairness.}

The goal of a system should be taken into account when deciding the methodology used for measuring bias\cite{rodolfafat2020}. This system prioritizes individuals for an assistive intervention to ensure they remain in medical care. A patient at risk for retention failure who does not receive an intervention loses an opportunity for their underlying challenges causing their risk to be addressed.  In this deployment setting where people are being flagged for an assistive intervention, bias is measured through metrics that measure disproportionate false negatives. Disproportionately failing to detect people of a certain group who are at risk for retention failure is more harmful than detecting false positives, where resources are wasted, because it can create a disparity between groups.  The system should, therefore, not disproportionately miss any at risk groups as this type of bias could exacerbate an already existing disparity or create new disparities. A false negative risk assessment carries less negative impact to the patient, though resource allocation can become more inefficient as interventions are wasted on patients who are falsely identified as high risk, leading to a trade-off between efficiency and fairness.

 An ideal model in production would be both efficient at reaching individuals (as captured by precision) and have minimal bias in missing individuals. To measure bias, we calculate the False Omission Rate (FOR), the ratio of the number of false negatives to the number of negative predictions, of protected groups. Given the racial composition of the population of PLWH in Chicago, we focused our attention on auditing models for parity in FOR by race. We considered a model to be disparate if its FOR ratio of Black vs White is less than 0.9 or greater than 1.1 (indicated by the purple band in the figures). Models were audited for bias using Aequitas \cite{saleiro_aequitas:_2018}, a python toolkit for auditing models for bias.

 \F\ref{fig:ucm_access_bias}, \ref{fig:ucm_retention_bias}, \ref{fig:cdph_bias} captures the criteria for an optimal model -- performance on the x-axis and bias on the y-axis. The purple band is the parity band with FOR ratio between 0.9 and 1.1. The dark lines are the 25th and 75th percentiles and the lighter lines are the minimum and maximum bias/performance observed, representing a measure of both stability of performance and bias. 

\subsection{HIV Clinic Bias Audit}
\subsubsection{HIV Clinic Retention in Care Bias Analysis}
\begin{figure}
	\includegraphics[width=\linewidth]{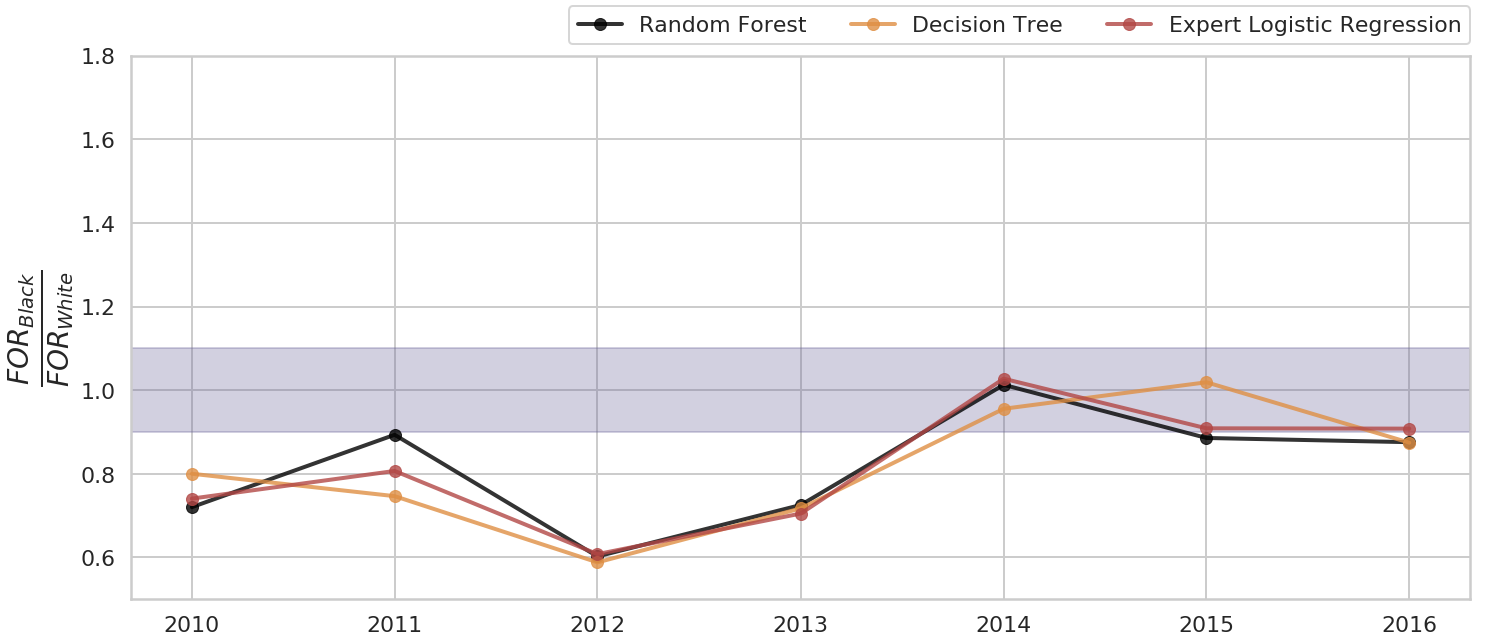}
	\includegraphics[width=\linewidth]{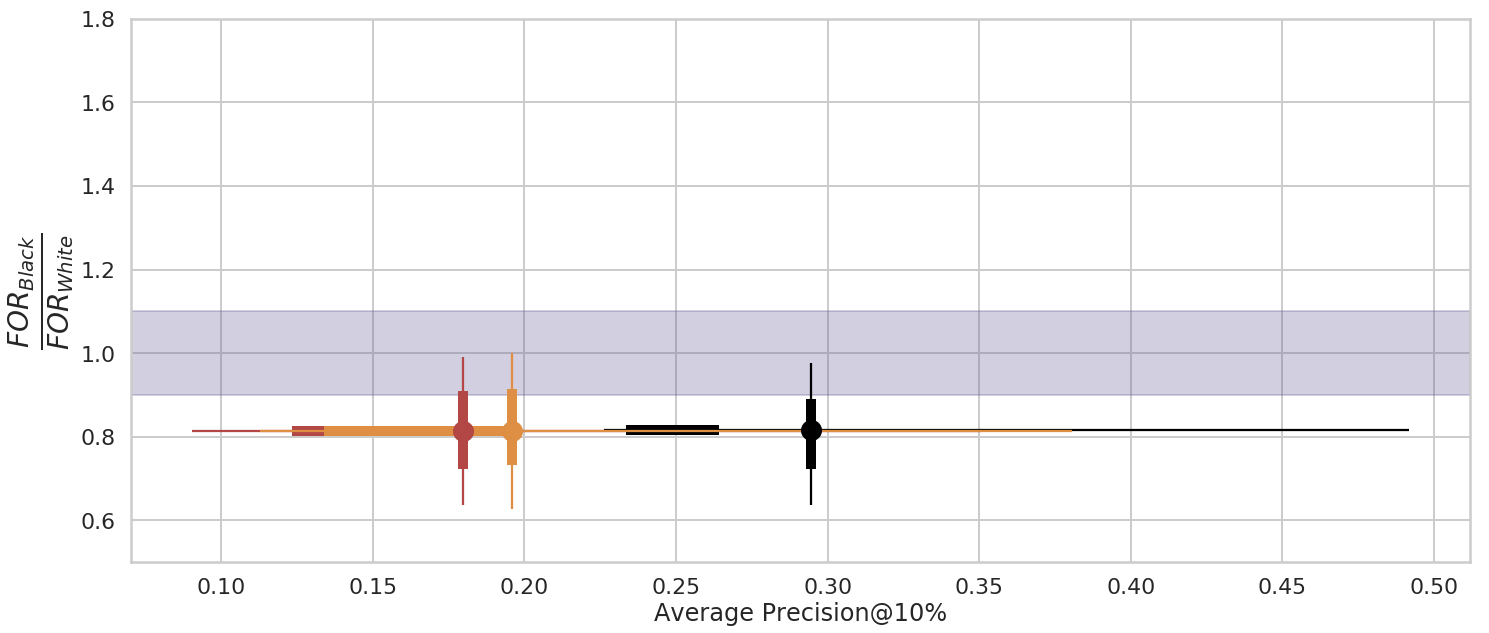}
    \caption{Bias in HIV Clinic Retention Models (Top) False Omission Rate among Black vs White in the model (black) compared to the logistic regression model (red) and Decision Tree model. Notably, the bias appears to be reducing in later time periods, FOR ratio is more often in the purple parity band in later years. (Bottom) FOR ratio compared to the precision@10\%. All the models are on average biased, but the random forest model has the best performance and the models have less bias in the later time periods. In this setting, the model selected for deployment is the random forest model since all other models have the same FOR ratio (bias) \textit{but} have worse performance.}
    \label{fig:ucm_retention_bias}
\end{figure}

The selected model for retention in care had FOR 0.26 $\pm$ 0.16 for black patients compared to 0.31 $\pm$ 0.17 for white patients (\F\ref{fig:ucm_retention_bias}). The expert logistic regression model had FOR of 0.27 $\pm$ 0.17 and 0.32 $\pm$ 0.17 for black and white patients respectively. Notably, as \F\ref{fig:ucm_retention_bias} shows the bias appears to be reducing in later time periods potentially indicating that the disparity is reducing with time. The FOR ratios for these models are all similar, indicating that the best model to choose would be the best performing model.

\subsubsection{HIV Clinic Access to Care Bias Analysis}
\begin{figure}
	\includegraphics[width=\linewidth]{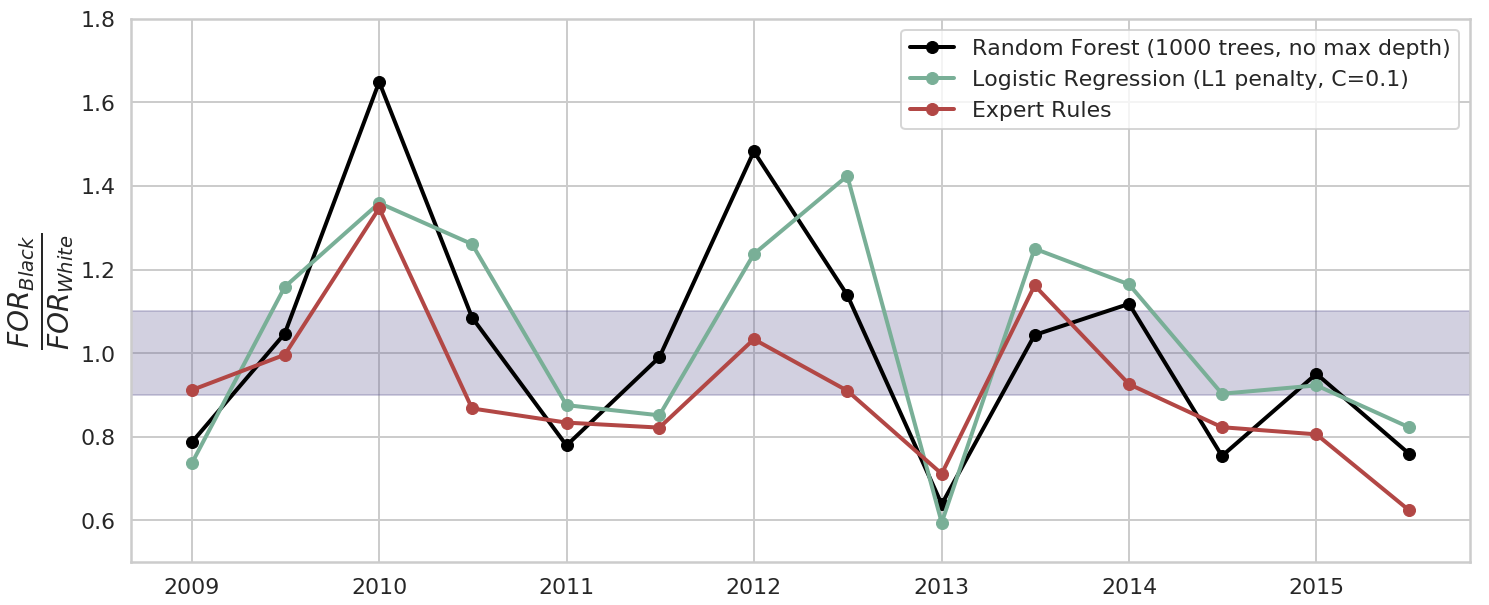}
	\includegraphics[width=\linewidth]{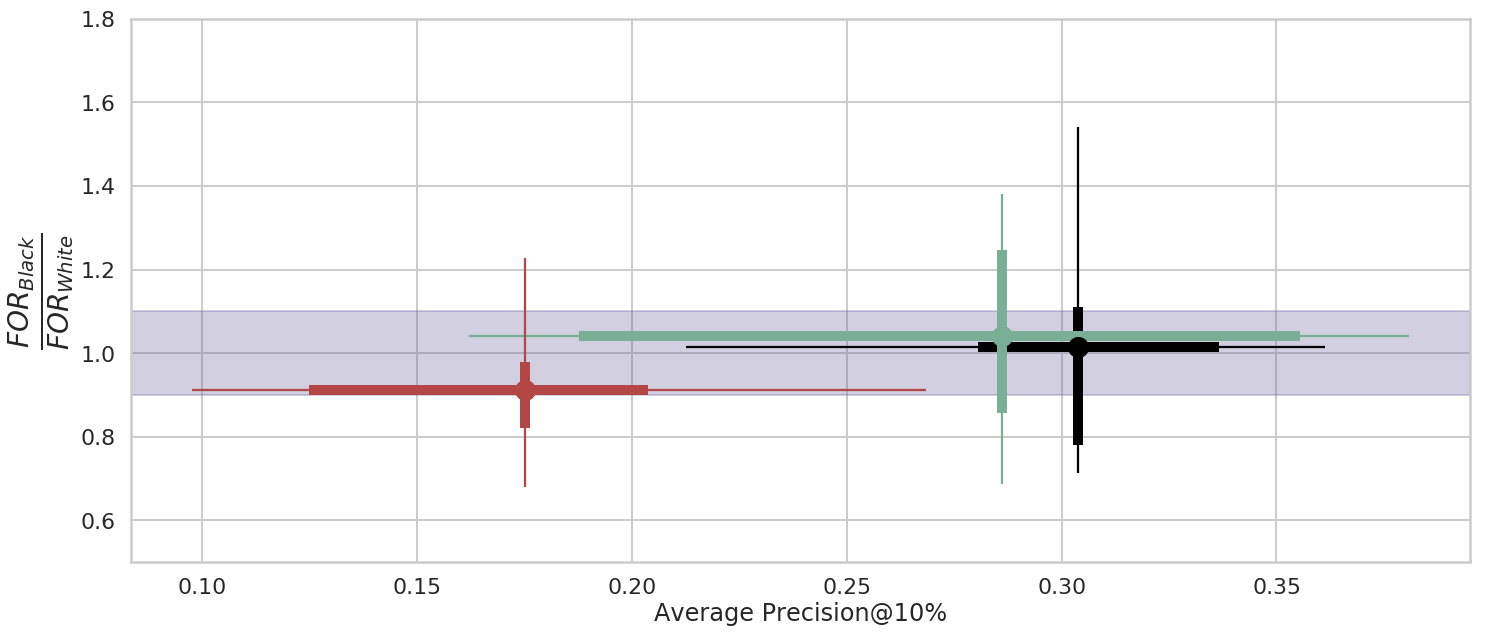}
    \caption{Bias in HIV Clinic Access Model (Top) False Omission Rate among Black vs White in the UCM access model (black) compared to the expert rules (red) and logistic regression (green) over time. While the Random Forest model has greater variance then the logistic regression or expert rules model, it falls within the parity band (purple) more often than the regression model. (Bottom) FOR for Black/African-American vs White compared to the precision@10\%. An ideal model will be farther right with higher precision and fall within the parity band. The Random Forest model's average FOR falls within the parity band and has a higher precision@10\% then the logistic regression model and expert rules, making it the chosen model}
    \label{fig:ucm_access_bias}
\end{figure}

The selected model for access to care in six months had FOR 0.24 $\pm$ 0.04 for black patients compared to 0.25 $\pm$ 0.08 for white patients (\F\ref{fig:ucm_access_bias}). The expert rules model had FOR of 0.26 $\pm$ 0.05 and 0.29 $\pm$ 0.08 for black and white patients respectively. It should be noted, the FOR ratios are calculated over a relatively small sample (120 appointments/year are flagged as high risk). As a result, this metric is susceptible to variation due to small population size. While the expert rules model has slightly less bias than the random forest, the trade-off in performance is high. This reduced performance as well as instability in the performance of the expert rules model make the Random Forest the model of choice.

\subsection{Health Department Bias Audit Results}

\begin{figure}[h]
    \includegraphics[width=\linewidth]{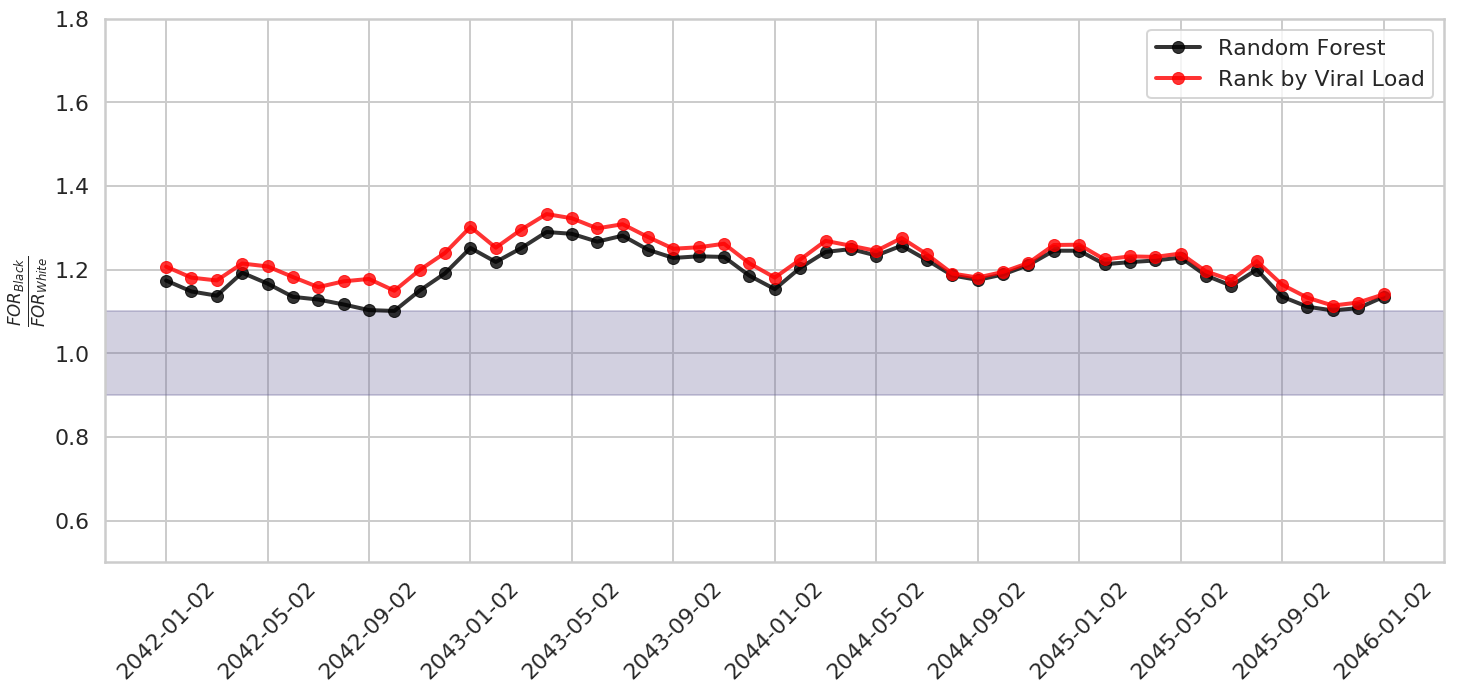}
	\includegraphics[width=\linewidth]{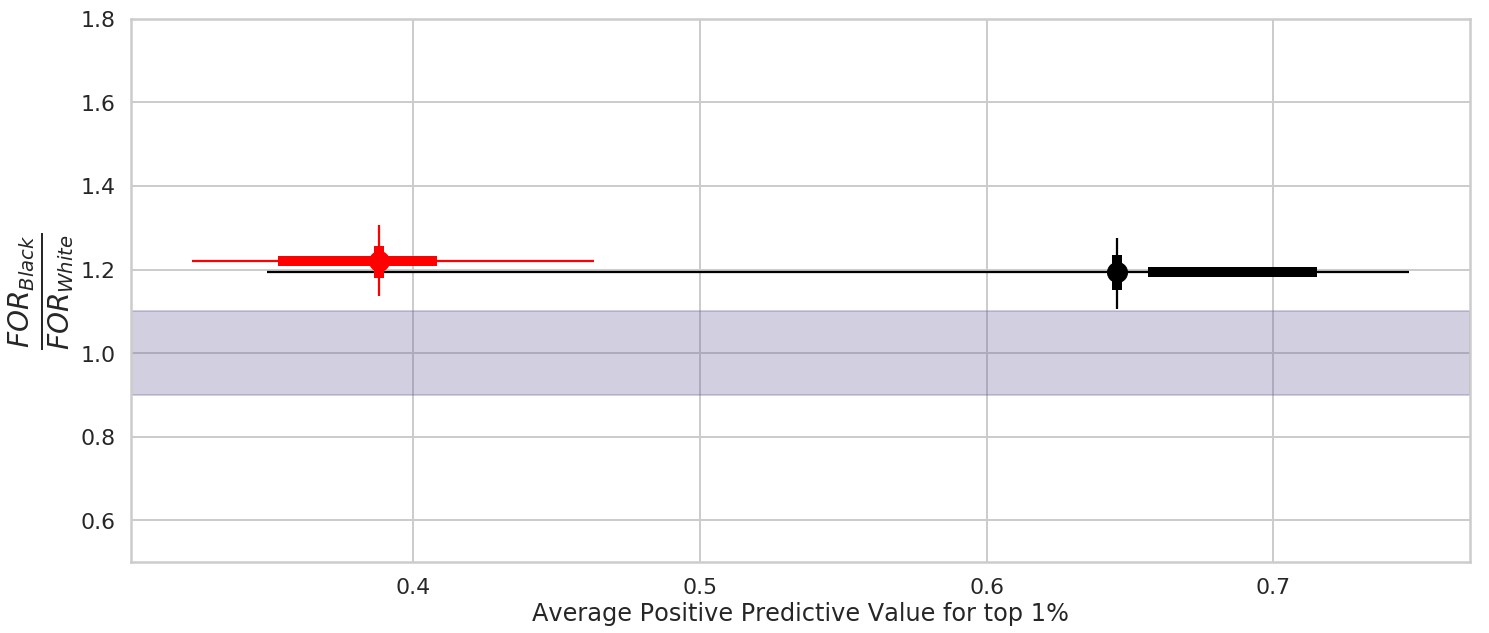}
    \caption{Bias in Public Health Retention model (Top) FOR ratio of Black/African-American vs White in our models compared to the ranking metric. (Bottom) FOR ratio vs precision@1\%. As with the HIV Clinic retention model, there is not a significant difference in bias among the models, indicating that the random forest that performs best based on precision@1\% is the best choice of model for deployment. Note that for all the models, the FOR ratio of Black vs White is above 1.2, indicating the model is disproportionately missing blacks more than whites.  }
    \label{fig:cdph_bias}
\end{figure}

The bias audit of the CDPH Access model found that the model was slightly biased towards missing more Black/African-American at-risk individuals (missing 20\% more) than white individuals but better than competing methods, including the expert rules. The selected model for access to care had FOR 0.30 $\pm$ 0.02 for black/African-American patients compared to FOR 0.24 $\pm$ 0.01 for white patients (\F\ref{fig:cdph_bias}). This bias should be monitored to understand if it is increasing or increasing disparity in retention outcomes before deployment. Models were also audited for bias across gender (male versus female) and risk category (MSM vs non-MSM) and did not find significant bias for those metrics. In addition, we tested across the model space and all models had similar or worse bias. For simplicity, we omitted these models from the figure. Given the similarity in bias across models, in this setting, the best choice would be selecting the best performing model and do further tests to ensure that it does not result in unfair outcomes.

In this setting, bias auditing is an important part of model selection; akin to selecting a model optimized for performance, a model can be optimized for fairness constraints to prevent disparity among protected groups.

\section{Impact and Conclusions}
This system demonstrates the potential of machine learning models to identify HIV patients that are at the highest risk of dropping out of medical care. The system can be used for point-of-care interventions in a clinic as well as proactive outreach by a public health department. The system has been specifically implemented for the University of Chicago HIV clinic and Chicago Department of Public Health.  Moreover, this methodology facilitates model selection based on performance under resource constraints, stability of performance over time, and fairness. While most prior work regarding retention in care examines factors associated with retention at a single point in time, our model dynamically predicts retention longitudinally. Patients’ appointment attendance patterns change over time, with patients often transitioning in and out of care \cite{lee_beyond_2018}. The system provides risk score at the visit level and recalculates the score as new data becomes available.

Bias in models can have an unexpected long term adverse impact of  on protected groups. To our knowledge, this is also the first use of bias auditing of predictive models in an HIV care setting.  We hope this work will engender further work to understand how to mitigate the risk of exacerbating disparities in more than just the HIV care setting.

Our methodology prioritizes \textit{all} the criteria of fairness, performance, and stability allowing for greater control of the real world impact of the model. It allows the clinic/public health department to balance the potentially competing goals. Other sites can replicate the process presented here for extracting electronic data and incorporating them into machine learning systems using the Triage framework~\cite{triage_dsapp} and our open source code\footnote{http://www.github.com/dssg}.     

\section*{Ethical Review of Study and Waiver of Consent}
\textit{University of Chicago HIV Clinic:}This UCM portion of this study was approved by the University of Chicago Institutional Review Board (IRB). The IRB waived the need for informed consent as part of the study approval. Research was carried out in accordance with the ethical standards in the Declaration of Helsinki.\textit{Chicago Department of Public Health:} The CDPH portion of this study was exempt from IRB approval.

\begin{acks}
The authors would like to acknowledge funding from the NIH-funded Third Coast Center for AIDS Research (CFAR) (P30 AI117943), Institute for Translational Medicine (UL1 TR000430), and National Institute of Health (K23MH121190-01).

\end{acks}

\bibliographystyle{ACM-Reference-Format}
\bibliography{hiv_bibliography}


\end{document}